\documentstyle[12pt]{article}
\newcommand{\non}{\nonumber}
\def\be{\begin{equation}}
\def\ee{\end{equation}}
\def\bea{\begin{eqnarray}}
\def\eea{\end{eqnarray}}

\def\a{\alpha}
\def\b{\beta}
\def\d{\delta}
\def\e{\epsilon}      

\def\k{\kappa}        
\def\l{\lambda}
\def\m{\mu}
\def\n{\nu}

\def\r{\rho}          

\def\x{\xi}

\def\F{\Phi}

\def\O{\Omega}

\def\ch{{\cal H}}

\def\cl{{\cal L}}

\def\pa{\partial}                             
\def\ha{\frac12}                              

\begin{document}

\title{{\bf An Extended Abelian Chern-Simons Model and the Symplectic Projector Method}}

\author{L.R.U. Manssur$ *, \dagger $, A.L.M.A. Nogueira$ *, \star$ and M.A. Santos$\ddagger$ \\
{$ *$\normalsize\it Centro Brasileiro de Pesquisas F\'{\i}sicas, CBPF - DCP}\\
{\normalsize\it Rua Dr. Xavier Sigaud 150, 22290-180 Rio de Janeiro, Brazil}\\
{$ \star$\normalsize\it Universidade Cat\'{o}lica de Petr\'{o}polis - Grupo
de F\'{\i}sica Te\'{o}rica, GFT - UCP}\\
{\normalsize\it Rua Bar\~{a}o do Amazonas 124, 25685-070 Petr\'{o}polis, Brazil}\\
{$\dagger$\normalsize\it Laborat\'{o}rio Nacional de Computa\c{c}\~{a}o Cient\'{\i}fica, LNCC-DMA}\\
{\normalsize\it Av. Get\'{u}lio Vargas 333, 25651-070 Petr\'{o}polis, Brazil}\\
{$\ddagger$\normalsize\it  Dep. de F\'{\i}sica, Univ. Fed. Rural do Rio de Janeiro (UFRRJ)}\\
{\normalsize\it 23851-180, Serop\'{e}dica, RJ-Brazil}}

\date{}

\maketitle

\vspace{-0.6cm}

\begin{abstract}
{\normalsize
\noindent 
The Symplectic Projector Method is applied to discuss quantisation aspects
of an extended Abelian model with a pair of gauge potentials coupled by
means of a mixed Chern-Simons term. We focuss on a field content that spans
an N=2-D=3 supersymmetric theory whenever scalar and fermionic matter is suitably coupled to the family of gauge potentials.}
\end{abstract}

\section{Introduction}

Nearly ten years ago, a method \cite{tudo} was developped that treats 
the fundamental question of canonically quantising field theories based 
on gauge symmetries. In this method, a crucial point is to identify, among 
the original constrained coordinates, those quantities related to the 
true degrees of freedom, which we refer to as the physical variables.
We shall call this procedure the Symplectic Projector Method (SPM).

Along this line of investigation, the SPM has been tested through a number 
of relevant situations, such as Classical Electrodynamics, the 
2-Dimensional Bosonised Schwinger Model \cite{2D}, the Christ-Lee Model \cite{CLee} and the Chern-Simons-Maxwell Theory \cite{CSM}.

In this work, we reassess the efficacy of the method in picking up the
true (physical) field coordinates for a sort of extended Abelian gauge 
model with a Chern-Simons term coupling a pair of gauge potentials. Such
a model is the 3D descent of a 4D gauge theory with a topological mass
term involving the Kalb-Ramond 2-form gauge potential. Also, the model
discussed here is the core of the (gauge) bosonic sector (we leave behind
an additive bilinear for a massive scalar field) of an $N=2$-supersymmetric
gauge theory that leads to (after suitable identifications of fields) an
$N=2$ model endowed with a rich structure of topological magnetic vortices \cite{no1, no2}.

Since our model displays a vector potential with a peculiar gauge 
transformation and without any dynamics, if taken on-shell, we believe
it could also provide an interesting working example to test the 
consistency and the efficacy of the SPM.

Our paper is presented according to the following outline: in Section 2,
we show explicitly the 4D origin of our model and perform its dimensional
reduction towards the 3D mixed Chern-Simons theory; in Section 3, we
establish the set of constraints and apply the SPM to pick up the 
physical variables. Finally, we present our General Conclusions. 
\section{The 4-Dimensional Model and Its Reduction}

The 4-dimensional- $U(1) \times U(1)$ model we start off is based
on the presence of a vector potential, $A_{\hat \m}$, together with
a rank-2 gauge potential, $B_{\hat \m \hat \n} = - B_{\hat \n \hat \m}$,
the latter playing the r\^{o}le of a Kalb-Ramond field  \cite{KR}. We use
hatted indices to denote components with respect to 4-dimensional space, 
while the bare ones refer to D=3 ($\hat{\mu} = 0,1,2,3 \;\; and \;\;
\mu = 0,1,2 $). The corresponding field strengths are given as below:
\bea
F_{\hat \m \hat \n} & \equiv & \pa_{\hat \m} A_{\hat \n} - \pa_{\hat \n} 
A_{\hat \m},  \\
G_{\hat \m \hat \n \hat \k} & \equiv & \pa_{\hat \m} B_{\hat \n \hat \k} 
+ \pa_{\hat \n} B_{\hat \k \hat \m} +  \pa_{\hat \k} B_{\hat \m \hat \n},
\eea
and the coupling with a general matter field is carried out by means of
the extended gauge-covariant derivative as below:
\be
D_{\hat \m} \F \equiv (\pa_{\hat \m} + i e A_{\hat \m} + i g G_{\hat \m}) \F,
\ee
where $G_{\hat \m}$ is the dual of the field strength 3-form:
\be
G_{\hat \m} \equiv \frac{1}{3!} \e_{\hat \m \hat \k \hat \l \hat \r}
G^{\hat \k \hat \l \hat \r}.
\ee
This means that charged matter couples minimally to $A_{\hat \m}$ (coupling
constant $e$) and non-minimally to $B_{\hat \m \hat \n}$ ($g$ is the coupling
parameter governing the non-minimal interaction).

We propose to begin with the 4D action as follows:
\be
\cl_{4D} = - \frac14 {F_{\hat \m \hat \n}}^2 + \frac{1}{12} {G_{\hat \m \hat 
\n \hat \k}}^2 + \ha m \e^{\hat \m \hat \n \hat \k \hat \l} A_{\hat \m}
\pa_{\hat \n} B_{\hat \k \hat \l} + (\mbox{matter-gauge terms});
\ee
we do not specify the matter-gauge coupling terms as we wish to discuss
the quantisation of the gauge potential sector exclusively. For a discussion
of the complete model, we refer the reader to the works of refs.
\cite{no1, no2}. $m$ is the mass parameter associated to the presence
of a massive spin-1 gauge boson in the spectrum \cite{winder}.

The idea is now to dimensionally reduce the model to (1+2) dimensions,
adopting the Scherk ansatz \cite{scherk}, by simply assuming that all 
fields do not depend on $x^3$:
\be
\pa_3(\mbox{fields}) = 0.
\ee
$A_{\hat\m}$ yields two independent fields in 3D, namely, a gauge potential,
$A_{\m}$, and a scalar, $\varphi \equiv A_3$; on the other hand, two vector
potentials stem from $B_{\hat{\m}\hat{\n}}$:
\be
B_\m \equiv B_{\m 3}
\ee
and
\be
Z_\m \equiv \ha \e_{\m\k\l} B^{\k\l}\; .
\label{Z}
\ee
The gauge transformations read now as given below:
\bea
A_\m^\prime & = & A_\m + \pa_\m \a, \label{transfs1} \\
B_\m^\prime & = & B_\m + \pa_\m \b, \label{transfs2} \\
Z_\m^\prime & = & Z_\m + \e_{\m\n\l} \pa^\n \x^\l, \label{transfs3}
\eea
where $\a, \b$ and $\x^\l$ are arbitrary functions: $\alpha $ is the gauge
parameter associated to the gauge symmetry of $A_{\hat{\mu}}$, whereas
$\beta $ and $\x^{\lambda}$ are the 3D descents ($\b \equiv \x_{3} $) of the 
(vector) gauge parameter associated to $B_{\hat{\mu}\hat{\nu}}$. The reduced model exhibits a $[U(1)]^3$-symmetry; the extra Abelian factor comes about
by virtue of the 4D gauge symmetry of the Kalb-Ramond field \cite{KR}.
Moreover, eq.(\ref{transfs3}) displays an unusual gauge transformation
for the vector potential $Z_{\mu}$, as we have anticipated. Such
an exchange of r\^{o}les between the longitudinal and transverse sectors
of the vector potential $Z_{\mu}$ (the longitudinal part is now 
gauge-invariant) has as a counterpart odd expressions for some of the
constraints, as we shall see in Section 3.

The relationships between the 4D and 3D field strengths are readily worked
out and the following expressions can be shown to hold:
\bea
F^2_{\hat \m \hat \n} & = & F^2_{\m \n} - 2 (\pa_\m \varphi)^2, \\
G^2_{\hat \m \hat \n \hat \k} & = & -3 G^2_{\m \n} + 6 (\pa_\m Z^\m)^2,\\
\e^{\hat \m \hat \n \hat \k \hat \l} A_{\hat \m} \pa_{\hat \n}
B_{\hat \k \hat \l} & = & 2 \e^{\m \n \k} A_{\m} \pa_{\n} 
B_{\k} - 2 \varphi (\pa_\m Z^\m),
\eea
so that the 3D Lagrangian for the gauge fields takes over the form:
\be
\cl ^{\mbox{gauge}}_{3D} = - \frac14 {F_{\m \n}}^2 + \ha 
(\pa_\m \varphi)^2 - \frac14 {G_{\m \n}}^2 + \ha (\pa_\m Z^\m)^2
+ m \e^{\m \n \k} A_{\m} \pa_{\n} B_{\k} - m \varphi (\pa_\m Z^\m),
\label{depart}
\ee
where $F_{\m \n}$ and $G_{\m \n}$ are the field strengths corresponding
to $A_\m$ and $B_\m$, respectively. One should also notice the presence
of a non-diagonal Chern-Simons term along with a partner that mixes
$ \varphi $ and $Z^\m$. The 3D action of eq.(\ref{depart}) is invariant
under the $U(1) \times U(1) \times U(1)$ -symmetry quoted in eqs. (\ref{transfs1}), (\ref{transfs2}) and (\ref{transfs3}).

In the work of ref. \cite{no1}, the potentials $A_\m$ and $B_\m$ were
suitably identified with each other in a consistent way in connection 
with an N=2-D=3 supersymmetric version of the Maxwell-Chern-Simons 
model with anomalous magnetic couplings of matter to the gauge fields.

Before going ahead to discuss the symplectic quantisation of the model
under consideration, we should perhaps mention that the 3 massive degrees
of freedom that propagate in 4D \cite{winder} are now accommodated in 
$A_\m$ (1 d.f.), $B_\m$ (1 d.f.) and $\varphi $ (1 d.f.); the vector
potential $Z_\m$, featured with the unusual gauge transformation given
in eq. (\ref{transfs3}) and with the potentially dangerous longitudinal
kinetic term ${(\partial_{\mu}Z^{\mu})}^{2}$, does not propagate any
on-shell degree of freedom (in fact, as displayed in eq.(\ref{Z}),
$Z_{\mu}$ is just the dual of the well-known non-propagating 3D
Kalb-Ramond field). This means that one of the $U(1)$-factors has no
dynamical significance.
The application of the Symplectic Projector Method to reassess the 
quantisation of this peculiar 3D gauge model in the sequel shall 
illustrate this procedure in a more evident way.
 
\section{Constraints and relevant degrees of freedom}

In view of what we have set previously, we define our model by means of
the Lagrangian density:
\be
\cl =-\frac{1}{4}F^{\mu \nu }F_{\mu \nu }-\frac{1}{4}G^{\mu \nu
}G_{\mu \nu }+\frac{1}{2}\partial ^{\mu }\varphi \partial _{\mu }\varphi +%
\frac{1}{2}\partial ^{\m }Z_{\m }\partial ^{\n }Z_{\n
}-m\left( \partial ^{\m }Z_{\m }\right) \varphi +m\varepsilon ^{\mu
\nu \rho }B_{\mu }\partial _{\nu }A_\rho ,
\label{3Dmodel}
\ee
with
\bea
F^{\mu \nu } & = & \partial ^\mu A^\nu -\partial ^\nu A^\mu , \\
G^{\mu \nu } & = & \partial ^\mu B^\nu -\partial ^\nu B^\mu , 
\eea
and the metric signature $\eta^{\mu\nu} = (+,-,-)$.
Picking up the canonically conjugate momenta, we have:
\begin{equation}
\pi ^\mu \equiv \frac{\delta \mathcal{L}}{\delta \left( \partial _0A_\mu
\right) }=-F^{0\mu }+m\varepsilon ^{\n 0\mu }B_\n \; ,
\end{equation}
which yields
\bea
\pi ^0 & = & 0 \\ 
\pi ^i & = & -F^{0i}+m\varepsilon ^{0ik}B_k.
\eea
Also,
\begin{equation}
P^\mu \equiv \frac{\delta \mathcal{L}}{\delta \left( \partial _0B_\mu \right) } = -G^{0\mu },
\end{equation}
or
\bea
P^0 & = & 0 \\ 
P^i & = & -G^{0i}.
\eea
For the scalar field,
\begin{equation}
\pi _{\varphi } \equiv \frac{\delta \mathcal{L}}{\delta \left( \partial
_{0}\varphi \right) }  =\partial _{0}\varphi.
\end{equation}
Finally,
\be
\pi^{\prime \mu} \equiv \frac{\delta \mathcal{L}}{\delta \left( \partial
_{0} Z_\m \right) } =  \left( -m\varphi +\partial ^\beta Z_\beta \right) \eta^{\mu 0}
\ee
or
\bea
\pi^{\prime 0} & = & \left( +\partial ^\beta Z_\beta - m\varphi \right) \\ 
\pi^{\prime i} & = & 0.
\eea

Now, we are ready to write down the canonical Hamiltonian of the theory:
\bea
\ch_c & = & \pi ^\mu \partial _0A_\mu +P^\mu \partial _0B_\mu +\pi ^{\prime
\mu} \partial _0Z_\mu +\pi _\varphi \partial _0\varphi -\cl  \non \\
 & = & \frac{1}{2}\pi _{i}^{2}+\frac{1}{2}P_{i}^{2}+\frac{1}{2}\pi
_{\varphi }^{2}+\frac{1}{2}\pi ^{\prime 2} _{0}+A_{0}\left( \partial _{i}\pi
_{i}\right) +B_0  \left( \partial _{i}P_{i}-m\varepsilon
_{0ij}\partial _{i}A_j   \right) +  \non \\
 & & +\frac{1}{4}F_{ij}F_{ij}+m\varepsilon _{0ik}\pi _{i}B_{k}+\frac{1}{2}
m^{2}B_{k}B_{k}+\frac{1}{4}G_{ij}G_{ij}+\frac{1}{2}\left( \partial
_{i}\varphi \right) ^{2}+\frac{1}{2}m^{2}\varphi ^{2} \non \\
 & & +\pi^\prime _{0}\left( m\varphi +\partial _{i}Z_{i}\right) \; .
\eea
The primary Hamiltonian is just
\be
\ch_p = \ch_c + v_i \O_i,
\ee 
where the primary constraints are
\bea
\Omega _{1} & = & \pi _{0}\approx 0 \non \\
\Omega _{2} & = & P_{0}\approx 0 \non \\
\Omega _{3} & = & \pi ^\prime _{1}\approx 0  \\
\Omega _{4} & = & \pi ^\prime _{2}\approx 0. \non
\eea
As one can immediately notice, $\Omega_{3}$ and $\Omega_{4}$ are the first counterparts of the $Z_{\mu}$'s unusual features to be brought about in the
set of constraints. The consistency condition imposed on the latter yields: 
\bea
\Omega _{5} & = & \partial _{i}\pi _{i}\approx 0 \non \\
\Omega _{6} & = & \partial _{i}P_{i}-m\varepsilon _{0ij}\partial _{i}A_j 
\approx 0 \\
\Omega _{7} & = & \pi ^\prime _{0}-f\left( t\right) \approx 0, \non 
\eea
for some arbitrary $f(t)$. They are all first class constraints; the gauge-fixing conditions will be so chosen that
\bea
\Omega _{8} & = & A_{0}\approx 0 \non \\
\Omega _{9} & = & B_{0}\approx 0 \non \\
\Omega _{10} & = &Z_{1}\approx 0 \non \\
\Omega _{11} & = & Z_{2}\approx 0 \\
\Omega _{12} & = & \partial _{i}A_{i}\approx 0 \non \\
\Omega _{13} & = & \partial _{i}B_{i}\approx 0 \non \\
\Omega _{14} & = & Z_{0}\approx 0, \non
\eea
where we have imposed the "Coulomb" gauge-fixing on $Z_{\mu}$ in straight analogy to the usual procedure as applied onto $A_{\mu}$ (and $B_{\mu}$).
As a net result, the collection of constraints related to $Z_{\mu}$ already
indicate that its phase space variables are to be excluded from the
dynamical subset. 

We have now to build the matrix 
$g_{ij}(x,y)=\left\{ \Omega _{i}(x),\Omega _{j}(y)\right\}$.
Adopting the notational convention 
$\delta \equiv \delta^{2} \left( x-y\right) $, we get:

\bea
\lefteqn{ g (x,y) = \hspace{2in}} \non \\
& & \non \\
& \!\!\! \!\!\!\!\!\! \!\!\!\! \left( 
\begin{array}{cccccccccccccc}
0 & 0 & 0 & 0 & 0 & 0 & 0 & -\delta & 0 & 0 & 0 & 0 & 0 & 0 \\ 
0 & 0 & 0 & 0 & 0 & 0 & 0 & 0 & -\delta & 0 & 0 & 0 & 0 & 0 \\ 
0 & 0 & 0 & 0 & 0 & 0 & 0 & 0 & 0 & \delta & 0 & 0 & 0 & 0 \\ 
0 & 0 & 0 & 0 & 0 & 0 & 0 & 0 & 0 & 0 & \delta & 0 & 0 & 0 \\ 
0 & 0 & 0 & 0 & 0 & 0 & 0 & 0 & 0 & 0 & 0 & \partial _{i}^{x}\partial
_{i}^{y}\delta & 0 & 0 \\ 
0 & 0 & 0 & 0 & 0 & 0 & 0 & 0 & 0 & 0 & 0 & 0 & \partial _{i}^{x}\partial
_{i}^{y}\delta & 0 \\ 
0 & 0 & 0 & 0 & 0 & 0 & 0 & 0 & 0 & 0 & 0 & 0 & 0 & -\delta \\ 
\hspace{.05in} \d \hspace{.05in} & 0 & 0 & 0 & 0 & 0 & 0 & 0 & 0 & 0 & 0 & 0 & 0 & 0 \\ 
0 & \hspace{.05in} \d \hspace{.05in} & 0 & 0 & 0 & 0 & 0 & 0 & 0 & 0 & 0 & 0 & 0 & 0 \\ 
0 & 0 & \hspace{.05in} -\d \hspace{.05in} & 0 & 0 & 0 & 0 & 0 & 0 & 0 & 0 & 0 & 0 & 0 \\ 
0 & 0 & 0 & \hspace{.05in} -\d \hspace{.05in} & 0 & 0 & 0 & 0 & 0 & 0 & 0 & 0 & 0 & 0 \\ 
0 & 0 & 0 & 0 & -\partial _{i}^{x}\partial _{i}^{y}\delta & 0 & 0 & 0 & 0 & 0
& 0 & 0 & 0 & 0 \\ 
0 & 0 & 0 & 0 & 0 & -\partial _{i}^{x}\partial _{i}^{y}\delta & 0 & 0 & 0 & 0
& 0 & 0 & 0 & 0 \\ 
0 & 0 & 0 & 0 & 0 & 0 & \hspace{.05in} \d \hspace{.05in} & 0 & 0 & 0 & 0 & 0 & 0 & 0
\end{array}
\right) \! \!  ,
\eea
whose inverse reads as below: 
\bea
\lefteqn{ g ^{-1} (x,y) = \hspace{2in}} \non \\
& & \non \\
& \!\!\! \!\!\!\!\!\! \!\!\! \left( 
\begin{array}{cccccccccccccc}
0 & 0 & 0 & 0 & 0 & 0 & 0 & \hspace{.05in} \d \hspace{.05in} & 0 & 0 & 0 & 0 & 0 & 0 \\ 
0 & 0 & 0 & 0 & 0 & 0 & 0 & 0 & \hspace{.05in} \d \hspace{.05in} & 0 & 0 & 0 & 0 & 0 \\ 
0 & 0 & 0 & 0 & 0 & 0 & 0 & 0 & 0 & \hspace{.05in} -\d \hspace{.05in} & 0 & 0 & 0 & 0 \\ 
0 & 0 & 0 & 0 & 0 & 0 & 0 & 0 & 0 & 0 & \hspace{.05in} -\d \hspace{.05in} & 0 & 0 & 0 \\ 
0 & 0 & 0 & 0 & 0 & 0 & 0 & 0 & 0 & 0 & 0 & +\nabla ^{-2} & 0 & 0 \\ 
0 & 0 & 0 & 0 & 0 & 0 & 0 & 0 & 0 & 0 & 0 & 0 & +\nabla ^{-2} & 0 \\ 
0 & 0 & 0 & 0 & 0 & 0 & 0 & 0 & 0 & 0 & 0 & 0 & 0 & \hspace{.05in} \d \hspace{.05in} \\ 
-\delta & 0 & 0 & 0 & 0 & 0 & 0 & 0 & 0 & 0 & 0 & 0 & 0 & 0 \\ 
0 & -\delta & 0 & 0 & 0 & 0 & 0 & 0 & 0 & 0 & 0 & 0 & 0 & 0 \\ 
0 & 0 & \delta & 0 & 0 & 0 & 0 & 0 & 0 & 0 & 0 & 0 & 0 & 0 \\ 
0 & 0 & 0 & \delta & 0 & 0 & 0 & 0 & 0 & 0 & 0 & 0 & 0 & 0 \\ 
0 & 0 & 0 & 0 & -\nabla ^{-2} & 0 & 0 & 0 & 0 & 0 & 0 & 0 & 0 & 0 \\ 
0 & 0 & 0 & 0 & 0 & -\nabla ^{-2} & 0 & 0 & 0 & 0 & 0 & 0 & 0 & 0 \\ 
0 & 0 & 0 & 0 & 0 & 0 & -\delta & 0 & 0 & 0 & 0 & 0 & 0 & 0
\end{array}
\right).
\eea

We shall label the fields and their corresponding momenta as follows:
\bea
\left( A^{0},A^{1},A^{2},\varphi ,B^{0},B^{1},B^{2},Z^{0},Z^{1},Z^{2},\pi
_{0},\pi _{1},\pi _{2},\pi _{\varphi },P_{0},P_{1},P_{2},\pi ^\prime _{0},\pi
^\prime _{1},\pi ^\prime _{2} \right) & \equiv& \non \\ 
\left( \xi _{1},\xi _{2},\xi _{3},\xi _{4},\xi _{5},\xi
_{6},\xi _{7},\xi _{8},\xi _{9},\xi _{10},\xi _{11},\xi
_{12},\xi _{13},\xi _{14},\xi _{15},\xi _{16},\xi
_{17},\xi _{18},\xi _{19},\xi _{20}\right). & & \non
\eea
At this stage, we are able to calculate the symplectic projector defined 
by the expression \cite{2D}:
\be
\Lambda _{j }^{i }\left( x,y\right) =\delta _{j }^{i }\delta
^{2}\left( x-y\right) -\varepsilon ^{ik}\int d^{2}z \; d^{2}w
  \left[ \delta _k (x) \Omega ^{m}\left(
z \right) \right]  g_{mn}^{-1}\left( z ,w \right) 
\left[ \delta _j (y)\Omega ^{n}\left( w \right) \right] \; ,
\ee
where
\begin{equation}
\delta _k (x)\Omega ^m\left( z \right) \equiv \frac{\delta
\Omega ^{m}\left( z \right) }{\delta \xi _{k }(x)},
\nonumber
\end{equation}
and $\varepsilon \equiv \left\{ \varepsilon^{ik} \right\} $ is the
symplectic matrix. The prescription to obtain the projected variables is
\be
\xi ^{i *}\left( x\right) =\int d^{2}y \; \Lambda _{j }^{i }\left(
x,y\right) \xi ^{j }\left( y\right),
\ee
yielding the following expressions for those which are non-trivial:
\bea
\xi ^{2*}\left( x\right) & = & A^{1\perp }\left( x\right)  \\
\xi ^{3*}\left( x\right)  & = & A^{2\perp }\left( x\right)  \\
\xi ^{4*}\left( x\right)  & = & \varphi \left( x\right)  \\
\xi ^{6*}\left( x\right)  & = & B^{1\perp }\left( x\right)  \\
\xi ^{7*}\left( x\right)  & = & B^{2\perp }\left( x\right)  \\
\xi ^{12*}\left( x\right)  & = & \pi _{1}^{\perp }\left( x\right) +m\int
d^{2}y\partial _{x_{2}}\nabla ^{-2}\left( x,y\right) \left[ \partial
_{y_{1}}B_{1}\left( y\right) +\partial _{y_{2}}B_{2}\left( y\right) \right]
  \non \\
 & = & \overline{\pi} _{1}^{\perp }\left( x\right)  \\
\xi ^{13*}\left( x\right)  & = & \pi _{2}^{\perp }\left( x\right) -m\int
d^{2}y\partial _{x_{1}}\nabla ^{-2}\left( x,y\right) \left[ \partial
_{y_{1}}B_{1}\left( y\right) +\partial _{y_{2}}B_{2}\left( y\right) \right]
 \non \\
 & = & \overline{\pi} _{2}^{\perp }\left( x\right) \\
\xi ^{14*}\left( x\right)  & = & \pi _{\varphi }\left( x\right) \\
\xi ^{16*}\left( x\right)  & = & P_{1}^{\perp }\left( x\right) +m\partial
_{x_{1}}\int d^{2}y\nabla ^{-2}\left( x,y\right) \left[ \partial
_{y_{1}}A_{2}\left( y\right) -\partial _{y_{2}}A_{1}\left( y\right) \right]
 \non \\
 & = & P_{1}^{\perp }\left( x\right) +m\partial _{x_{1}}\int d^{2}y\nabla
^{-2}\left( x,y\right) \left[ \mathbf{\nabla \times }A^{\perp }\right] _{y} \\
\xi ^{17*}\left( x\right)  & = & P_{2}^{\perp }\left( x\right) +m\partial
_{x2}\int d^{2}y\nabla ^{-2}\left( x,y\right) \left[ \partial
_{y_{1}}A_{2}\left( y\right) -\partial _{y_{2}}A_{1}\left( y\right) \right]
 \non \\ 
 & = & P_{2}^{\perp }\left( x\right) +m\partial _{x_{2}}\int d^{2}y\nabla
^{-2}\left( x,y\right) \left[ \mathbf{\nabla \times }A^{\perp }\right] _{y} 
\; .
\eea

The reduced canonical Hamiltonian is obtained from the previous
primary Hamiltonian taking into account the constraints and the gauge
conditions, now looked upon as strong equalities. We obtain:
\bea
\ch_{c}^{r} & = & \frac{1}{2}\pi _{i}^{2}+\frac{1}{2}P_{i}^{2}+\frac{1}{2}\pi
_{\varphi }^{2}+\frac{1}{2}\pi ^{\prime 2} _{0}+\frac{1}{4}%
F_{ij}F_{ij}+m\varepsilon _{0ik}\pi _{i}B_{k}+\frac{1}{2}m^{2}B_{k}B_{k}+%
 \non \\
 & & +\frac{1}{4}G_{ij}G_{ij}+\frac{1}{2}\left( \partial _{i}\varphi \right) ^{2}+\frac{1}{2}%
m^{2}\varphi ^{2}+\pi ^\prime _{0}\left( m\varphi \right), 
\eea
or, in symplectic notation,
\bea
\ch_{c}^{r} & = & \frac{1}{2}\left( \xi _{12}^{2}+\xi _{13}^{2}\right) +\frac{%
1}{2}\left( \xi _{16}^{2}+\xi _{17}^{2}\right) +\frac{1}{2}\xi _{14}^{2}+%
\frac{1}{2}\xi _{18}^{2}+\frac{1}{2}\left( \partial _{1}\xi _{3}\right) ^{2}
+ \frac{1}{2}\left( \partial _{2}\xi _{2}\right) ^{2}+ \non \\
 & & -\left( \partial _{1}\xi _{3}\right) \left( \partial _{2}\xi _{2}\right)
-m\left( \xi _{12}\xi _{7}-\xi _{13}\xi _{6}\right) +\frac{1}{2}m^{2}\left(
\xi _{6}^{2}+\xi _{7}^{2}\right) +\frac{1}{2}\left( \partial _{1}\xi
_{7}\right) ^{2}+\frac{1}{2}\left( \partial _{2}\xi _{6}\right) ^{2}+ \non \\
 & & -\left( \partial _{1}\xi _{7}\right) \left( \partial _{2}\xi _{6}\right) +
\frac{1}{2}\left( \partial _{1}\xi _{4}\right) ^{2}+\frac{1}{2}\left(
\partial _{2}\xi _{4}\right) ^{2}+\frac{1}{2}m^{2}\xi _{4}^{2}+\xi
_{18}\left( m\xi _{4} \right).
\eea
The physical Hamiltonian density is obtained by rewriting the one given above
in terms of the projected variables:
\bea
\ch^{*} & = & \frac{1}{2}\left( \xi _{12}^{*2}+\xi _{13}^{*2}\right)
+\frac{1}{2}\left( \xi _{16}^{*2}+\xi _{17}^{*2}\right) +\frac{1}{2}\xi
_{14}^{*2}+\frac{1}{2}\left( \partial _{1}\xi _{3}^{*}\right) ^{2}+\frac{1}{2%
}\left( \partial _{2}\xi _{2}^{*}\right) ^{2}+ \non \\
 &  & -\left( \partial _{1}\xi _{3}^{*}\right) \left( \partial _{2}\xi
_{2}^{*}\right) -m\left( \xi _{12}^{*}\xi _{7}^{*}-\xi _{13}^{*}\xi
_{6}^{*}\right) +\frac{1}{2}m^{2}\left( \xi _{6}^{*2}+\xi _{7}^{*2}\right) +%
\frac{1}{2}\left( \partial _{1}\xi _{7}^{*}\right) ^{2}+ \non \\
 & & +\frac{1}{2}\left( \partial _{2}\xi _{6}^{*}\right) ^{2} -\left( \partial
_{1}\xi _{7}^{*}\right) \left( \partial _{2}\xi _{6}^{*}\right) +\frac{1}{2}%
\left( \partial _{1}\xi _{4}^{*}\right) ^{2}+\frac{1}{2}\left( \partial
_{2}\xi _{4}^{*}\right) ^{2}+\frac{1}{2}m^{2}\xi _{4}^{*2} \; .
\eea
Finally, the equations of motion are obtained directly from the Hamilton-Jacobi equations by means of Poisson parentheses of the projected
variables with the Hamiltonian $ \int d^{2} y \;\ch^{*}(y) $. In so doing, we arrive at:
\be
{\stackrel{.}{\xi }}_{4}^{*} (x)  = \int d^{2}y \left\{ \xi _{4}^{*}(x),\ch
^{*}(y)\right\} = \xi _{14}^{*}(x) \; ;
\ee
this yields
\be
{\stackrel{..}{\xi }}_{4}^{*} = {\stackrel{.}{\xi }}_{14}^{*} = 
\int d^{2}y\left\{ \xi _{14}^{*},\ch^{*}(y)\right\}=-m^{2}\xi _{4}^{*}+\nabla ^{2}\xi _{4}^{*}
\; ,
\nonumber
\ee
or
\begin{equation}
(\Box + m^{2})\xi _{4}^{*} = 0 \; .
\end{equation}
Analogously, we obtain
\bea
{\stackrel{..}{\xi }}_{2}^{*} & = & \partial _{2}\partial _{2}\xi _{2}^{*}-\partial
_{1}\partial _{2}\xi _{3}^{*}-m\xi _{17}^{*} \non \\ 
{\stackrel{..}{\xi }}_{3}^{*} & = & \partial _{1}\partial _{1}\xi _{3}^{*}-\partial
_{1}\partial _{2}\xi _{2}^{*}+m\xi _{16}^{*}  \non \\
{\stackrel{..}{\xi }}_{6}^{*} & = & -m^{2}\xi _{6}^{*}+\partial _{2}\partial _{2}\xi
_{6}^{*}-\partial _{1}\partial _{2}\xi _{7}^{*}-m\xi _{13}^{*} \label{fe} \\
{\stackrel{..}{\xi }}_{7}^{*} & = & -m^{2}\xi _{7}^{*}+\partial _{1}\partial _{1}\xi
_{7}^{*}-\partial _{1}\partial _{2}\xi _{6}^{*}+m\xi _{12}^{*}. \non
\eea
Now, eqs.(\ref{fe}) can be rephrased in a much simpler form if one chooses, without loss of generality, the momentum to lay upon the x-axis
($\vec{k} = (k,0)$), selecting the components ($A_{2}, B_{2}$) to be the transverse ones. One can easily notice that such a choice would cancel the variables $\xi^{*}_{2}$, $\xi^{*}_{6}$ and $\xi^{*}_{12}$, and render the variables $\xi^{*}_{16}$ and $\xi^{*}_{17}$ equal to:
\bea
\xi^{*}_{16} & = & -m \xi^{*}_{3} ,\non \\
\xi^{*}_{17} & = & P_{2}^{\perp } .\non
\eea

The set of independent variables would correspondingly be specified by
the pairs ($\xi^{*}_{3}, \xi^{*}_{13}$), ($\xi^{*}_{4}, \xi^{*}_{14}$) and ($\xi^{*}_{7}, \xi^{*}_{17}$). The remaining equations of motion would then read:

\bea
\Box \xi_{4}^{*} & = & - m^{2}\xi _{4}^{*} , \non \\
\Box \xi_{3}^{*} & = & - m^{2}\xi _{3}^{*} , \non \\
\Box \xi_{7}^{*} & = & - m^{2}\xi _{7}^{*} . 
\label{modes}
\eea

So, we have a massive scalar, $\x_4^*$, and two massive transverse vector
fields, $\x_3^* $ and $\x_7^* $, according to what could be expected
from our counting of degrees of freedom in the framework of the 3D
Lagrangian given by eq.(\ref{3Dmodel}), if we were to keep the mapping $B_{\mu\nu}\rightarrow Z_{\mu} $ in mind. Such an outcome also matches
the natural allocation of physical degrees of freedom that could be
inferred from the original 4D model. Moreover, the fact that the two
vectors provide the physical sector with equivalent transverse massive contributions indicates the room for a consistent mapping into a new
model, a procedure that can be implemented through the identification
of the vector fields (and partners, in a supersymmetric context), as
performed in Ref.\cite{no1}.

\section{General Conclusions}
We have reassessed the efficacy of the SPM in selecting the true dynamical
set of phase space variables for a gauge 3D model hosting a peculiar
topological term, namely, a {\it mixed} Chern-Simons bilinear. As a 
consequence of the dimensional reduction procedure, which defines our
model as a descent of the 4D Cremmer-Scherk-Kalb-Ramond model
\cite{KR,Crem}, a counterpart for the mixed Chern-Simons term shows up
as a source of mass for the scalar field. The SPM has proven to be
efficient in casting the physical variables and exhibiting their dynamics through the field equations (\ref{modes}), where the expected topological
mass generation is explicitly displayed.

Concerning the results obtained through Dirac's \cite{Dirac} method for
the canonical quantisation of gauge systems, as applied to our 3D model,
the dynamics turns out to be the same. This convergence of outcomes
can be seen as another check of consistency for the SPM. Moreover, in
contrast to the possibility, {\it in principle}, to reduce the phase
space through Dirac's approach, by using the SPM we always place ourselves
in the right context for totally identifying, with the very explicit corresponding expressions, the true physical variables, an outcome lying 
on the very heart of the geometrical nature of the method.

The analysis we have pursued in this work happens to be an achievement in
the sense of fully and rigorously specifying the dynamics generated
in the gauge sector of an interesting N=2 off-shell supersymmetric model,
one that can be mapped (through suitable identifications of fields) into
another N=2 system in which topological self-dual vortex solutions can be
found on-shell \cite{no1,no2}. The presence of two massive transverse
vector excitations along with a massive scalar mode in the (bosonic) gauge sector is thus the relevant information confirmed as part of our knowledge
about the above mentioned N=2-D=3 model. Also, applying the Symplectic
Projector Method has proven to be a consistent choice for the complete clarification of the symplectic structure underlying the phase space
spanned by this gauge model.

\section*{Acknowledgements}

The authors are grateful to Prof. J.A. Helay\"{e}l-Neto for useful 
suggestions, discussions and the always kindest support to our
work. L.R.U. Manssur and M.A. Santos also express their gratitude to
the GFT-UCP (Group of Theoretical Physics),
at Universidade Cat\'{o}lica  de Petr\'{o}polis - UCP 
(Petr\'{o}polis, Brazil), where part of this work has been done,
for the kind and warm hospitality.

\vspace{0.5cm}

\noindent{\bf E-mail contact}:

\noindent{$\dagger$\normalsize\it leon@cbpf.br, {$\star$} nogue@cbpf.br}

\end{document}